\documentclass[conference]{IEEEtran}
\usepackage{amsmath,amssymb,amsfonts}
\usepackage{graphicx}
\usepackage{textcomp}
\usepackage{xcolor}
\usepackage{cases}
\usepackage{algorithm}
\usepackage[noend]{algpseudocode}
\usepackage{tikz} 
\usepackage{cite}
\usepackage{comment}
\usepackage{balance}

\newtheorem{remark}{Remark}
\newtheorem{theorem}{Theorem}

\newtheorem{lemma}{Lemma} 

\newtheorem{example}{Example}

\newtheorem{assumption}{Assumption}

\newcommand{\bs}{\boldsymbol}

\newcommand\independent{\protect\mathpalette{\protect\independenT}{\perp}}
\def\independenT#1#2{\mathrel{\rlap{$#1#2$}\mkern2mu{#1#2}}}

\renewcommand{\thepage}{\arabic{page}}
\renewcommand{\thesection}{\arabic{section}}
\renewcommand{\theequation}{\arabic{equation}}
\renewcommand{\thefigure}{\arabic{figure}}
\renewcommand{\thetable}{\arabic{table}}
\newcommand{\argmin}{\operatornamewithlimits{argmin}}

\usepackage{xspace}
\usepackage{bbm}
\usepackage{mathrsfs}

\def\Var{\mathop{\rm Var}\nolimits}%
\def\Cov{\mathop{\rm Cov}\nolimits}%
\def\e{\epsilon}

\def\l{\lambda}

\DeclareMathOperator\E{\sf E}

\newcommand{\U}{\mathrm{Unif}}

\def\textiid{i.i.d.\@\xspace}
\newcommand\iid{\ifmmode\text{ i.i.d. } \else \textiid \fi}

\begin{document}
\title{An Invariant Matching Property for Distribution Generalization under Intervened Response 
}

\author{%
  \IEEEauthorblockN{Kang Du and Yu Xiang}
  \IEEEauthorblockA{University of Utah\\
                    50 S Central Campus Dr \#2110 Salt Lake City, UT, USA\\ 
                    Email: \{kang.du, yu.xiang\}@utah.edu}
}

\maketitle

\begin{abstract}
The task of distribution generalization concerns making reliable prediction of a response in unseen environments. The structural causal models are shown to be useful to model distribution changes through intervention. Motivated by the fundamental invariance principle, it is often assumed that the conditional distribution of the response given its predictors remains the same across environments. However, this assumption might be violated in practical settings when the response is intervened. In this work, we investigate a class of model with an intervened response. We identify a novel form of invariance by incorporating the estimates of certain features as additional predictors. Effectively, we show this invariance is equivalent to having a deterministic linear matching that makes the generalization possible. We provide an explicit characterization of the linear matching and  present our simulation results under various intervention settings.   
\end{abstract}

\begin{IEEEkeywords}
Distribution generalization, invariance, causal structural model, LMMSE estimator
\end{IEEEkeywords}


\allowdisplaybreaks
\section{Introduction}
Consider the problem of predicting the response $Y$ given its predictors $X = (X_1,...,X_d)^\top$ in unseen environments. 
To model distribution changes in different environments (or training and test distributions), the common assumption is that the assignment for $Y$ does not change across environments (or $Y$ is not intervened). The structural causal models (SCMs)~\cite{pearl2009causality,peters2017elements} allow for natural formulations of the conditional distribution of $Y$ given $X$~\cite{scholkopf2012causal,zhang2013domain,peters2016causal,peters2017elements,rojas2018invariant,heinze2017conditional,buhlmann2020invariance}, and the underlying principle is known as invariance, autonomy or modularity~\cite{haavelmo1944probability,aldrich1989autonomy,pearl2009causality,imbens2015causal}. 
For instance, in the invariance causal prediction framework~\cite{peters2016causal}, it is assumed that the conditional distribution of $Y$ given a set of predictors $X_S\subseteq\{X_1,...,X_d\}$ is invariant in all environments; a relaxed version is adopted in the stabilized regression method~\cite{pfister2021stabilizing} where only the conditional mean is assumed to be invariant. 

In practical settings, however, the structural assignment of $Y$ might change across environments, i.e., $Y$ might be intervened. We thus believe there is a need of relaxing the assumption and exploring alternative forms of invariance. To shed some light on this more challenging setting, we propose to model $Y$ as  
\begin{align*}
	Y = f_U(X_{\text{PA}(Y)}, \e_Y),
\end{align*}
where $\text{PA}(Y)$ denotes the set of direct (causal) parents of $Y$, and $\e_Y$ is an independent noise, and a (discrete) random variable $U$ is introduced to capture the dependence of structural assignment on different environments (i.e., each $U=u$ corresponds to one environment). The main challenge lies in whether it is still possible to identify forms of invariance to facilitate prediction in unseen environments. In this work, we make an attempt in this direction by focusing on a mixture of linear SCM models, and the assignment for $Y$ is
\begin{align}
		Y = a(U)^\top X_{\text{PA}(Y)}^1+ b^\top X_{\text{PA}(Y)}^2   + \e_Y,\label{Y_assign}
\end{align}
where $X_{\text{PA}(Y)}$ is partitioned into $X_{\text{PA}(Y)}^1$ and  $X_{\text{PA}(Y)}^2$, and coefficients $a(U)$ formalize the changing conditional distributions; furthermore, we consider a training model and a testing model, and allow $a(\cdot)$ to be arbitrarily different under the two models. To make reliable predictions on the testing model, we identify an additional class of predictors that are computed based on the linear minimum mean square error (LMMSE) estimators of $X_k$ given $X_S$ for any fixed $U=u$, for $k\in \{1,...,d\}$ and $S\subseteq\{1,...,d\} \setminus k$. Roughly speaking, these predictors (along with the original ones $X$) allows for a deterministic relation for predicting $Y$, with coefficients that are invariant for all environments. This makes the generalization task possible, as one can then reuse the coefficients for unseen environments or test data. 

\section{Background and Problem Formulation} 

We now formally introduce the model and rewrite~\eqref{Y_assign} in a more compact form. Let $\mathcal{U}$ denote a set of environments where a target variable $Y \in \mathbb{R}$ and a vector of predictors  $X=(X_{1},\ldots,X_{d})^{\top} \in \mathcal{X}\subseteq \mathbb{R}^{d \times 1}$ are observed, we assume $U \in \mathcal{U}$ and $(X,Y)$ satisfy an acyclic SCM that we call the training SCM,
 \begin{numcases}{\mathcal{S}:}
    U = \varepsilon_{U} \nonumber\\
    X = \gamma Y + BX+\varepsilon_{X}\label{scm_x}\\
    Y = (\beta +  \alpha(U))^{\top}X +  \varepsilon_{Y}, \label{scm_y}
\end{numcases}
where $\varepsilon_{U} \in \mathcal{U}$, $\varepsilon_{Y}\in \mathbb{R}$, $\alpha(U), \beta, \gamma, \varepsilon_{X} \in \mathbb{R}^{d \times 1}$, $B \in \mathbb{R}^{d \times d}$, and the noise variables $\varepsilon_{X_1},\ldots , \varepsilon_{X_d}, \varepsilon_{U}$, and $\varepsilon_{Y}$ are jointly independent, and assume that $\alpha(U)$ is a nondegenerate random variable (i.e., not a one-point distribution). This nonlinear SCM $\mathcal{S}$ can be viewed as a mixture of linear SCMs, since $\mathcal{S}$ is linear when conditioning on $U=u$. The causal graph $\mathcal{G}(\mathcal{S})$ induced by $\mathcal{S}$ can be drawn according to the nonzero coefficients in $\mathcal{S}$. Without loss of generality, we require $\{X_j: \beta_{j}\neq 0 \}$ and $\{X_j:\alpha_{j}(U)\neq 0 \}$ to be two distinct sets of parents of $Y$. 

We assume that the variable $U$ is a root node in $\mathcal{G}(\mathcal{S})$ such that only the parameters that are functions of $U$ may change in the testing SCM defined below. The reason behind this assumption is that if there is no evidence that a parameter is changing for a diverse set of environments in the training data, then that parameter is likely to remain invariant in the test data or any unseen environments. 
\smallskip
\begin{remark}
For an intercept term in~\eqref{scm_y} that depends on $U$, it can be taken as the coefficient of $X_1=1$. For simplicity, we assume that $\varepsilon_{X}$ and $\varepsilon_{Y}$ have zero means, which implies $\E[X|U=u]=\E[Y|U=u]=0$ and thus $\E[X]=\E[Y]=0$ (the same goes for the testing SCM defined below).
\end{remark}

\begin{remark}
In~\cite{du2020causal}, the authors have shown that a form of varying filter connecting feature and response (as a special case of the varying coefficients in~\eqref{scm_y}) is effective for causal inference tasks, by adopting estimators from~\cite{xiang2019estimation}.
\end{remark}
\smallskip


Similarly, let $\mathcal{U}^{\tau}$ denote a set of unseen environments ($U^{\tau} = u^{\tau}$) where the observed variables $U^{\tau} \in \mathcal{U}^{\tau}$ and $X^{\tau}=(X^{\tau}_{1},\ldots,X^{\tau}_{d})^{\top} \in \mathcal{X}^{\tau} \subseteq \mathbb{R}^{d \times 1}$ and the unobserved variable $Y^{\tau} \in \mathbb{R}$ follow an acyclic testing SCM,
\begin{align}
\mathcal{S}^{\tau}:\label{scm_s_testing}
 \begin{cases}
    U^{\tau} = \varepsilon_{U^{\tau}}\\
    X^{\tau} = \gamma Y^{\tau} + BX^{\tau} +  \varepsilon_{X^{\tau}} \nonumber \\
    Y^{\tau} = (\beta +  \alpha^{\tau}(U^{\tau}))^{\top}X^{\tau} +\varepsilon_{Y^{\tau}},
\end{cases}
\end{align}
where the noise variables $\varepsilon_{X^{\tau}_1},\ldots, \varepsilon_{X^{\tau}_d}, \varepsilon_{U^{\tau}}$,  and $\varepsilon_{Y^{\tau}}$ are jointly independent, and $(\varepsilon_{X^{\tau}}^{\top},\varepsilon_{Y^{\tau}})$ and  $(\varepsilon^{\top}_{X},\varepsilon_{Y})$ are equal in distribution. Since we assume that only the parameters that are functions of $U$ in $\mathcal{S}$ may change in $\mathcal{S}^{\tau}$, we have $\alpha_{j}^{\tau}(U^{\tau})=0$  for any $j \in \{1,\ldots,d\}$ such that $\alpha_{j}(U)=0$. 

In this work, we consider the setting when the only parameter in $\mathcal{S}$ that depends on $U$ is the coefficient vector $\alpha(\cdot)$, but $\alpha(\cdot)$ and $\alpha^{\tau}(\cdot)$ can be arbitrarily different. By assuming the independence of $\varepsilon_{U}$ and $\varepsilon_{Y}$, the distribution of $\varepsilon_{Y}$ remains invariant when conditioning on $U=u$ for different $u$, while the case when the variance of $\varepsilon_{Y}$ changes arbitrarily with respect to $u$ can be challenging, since $\Var(\varepsilon_{Y})$ is simply the MMSE of the estimator $\E[Y|X_{\text{PA}(Y)},U]$. Another setting when only the parameters in the assignments of the predictors are allowed to depend $U$ (i.e., only $X$ is intervened) is considered in the stabilized regression framework~\cite{pfister2021stabilizing}, where a weaker version of the causal invariance property~\cite{peters2016causal} is assumed. Using our notation, it is assumed that there exists $S \subseteq \{1,\ldots,d\}$ such that $\E[Y|X_{S}=x, U=u] = \E[Y|X_{S}=x] \triangleq g(x)$
holds for all $x$ and $u$. Since $g(x)$ does not depend on $u$, the above relation remains the same for both the training and testing SCMs (for instance, $\E[Y^{\tau}|X^{\tau}_{S}=x^{\tau}, U^{\tau}=u^{\tau}] = \E[Y^{\tau}|X^{\tau}_{S}=x^{\tau}] = g(x^{\tau}) $ holds for all $x$ and $u^{\tau}$). This assumption allows one to select predictors that provide consistent predictions of the target variable across the observed and unseen environments. In general, the assumption is violated for the SCMs $\mathcal{S}$ and $\mathcal{S}^{\tau}$ when $Y$ is intervened, or equivalently, when the parameters depending on $U$ appear in the assignment of $Y$. 


Our invariance property relies on the LMMSE estimators of a target variable $Y \in \mathbb{R}$ given a vector of predictors $X \in \mathbb{R}^{p \times 1}$, denoted by $\E_{l}[Y|X] =(\theta^{\text{ols}})^{\top}(X-\E[X]) + \E[Y]$,
where $\theta^{\text{ols}} \triangleq \Cov(X,X)^{-1} \Cov(X,Y)$ is also called the population ordinary least squares (OLS) estimator. For the SCM $\mathcal{S}$, we denote the LMMSE estimator of $Y$ given $X$ when conditioning on $U=u$ as $ \E_{l}[Y|X; U= u ] \triangleq (\theta^{\text{ols}}(u))^{\top}X$ with its OLS estimator $\theta^{\text{ols}}(u) \in \mathbb{R}^{d \times 1}$. And correspondingly, we define $\E_{l}[Y|X; U] \triangleq (\theta^{\text{ols}}(U))^\top X$ that is linear in $X$ but with coefficients depending on $U$. Equivalently, one can define $\E_{l}[Y|X; U]$ by
\begin{equation}
\E_{l}[Y|X; U] =   \argmin_{l^{\top}(U)X \in \mathcal{L} } \E\left[|Y-l^{\top}(U)X|^2\right],  \label{eq:lmmse_U}
\end{equation}
where $\mathcal{L} = \{l^{\top}(U)X\,|\, l:\mathcal{U} \to \mathbb{R}^{d \times 1}\}$ is a class of functions that are linear in $X$ but with coefficients depending on $U$. This function class is introduced as it is compatible with the form of the assignment of $Y$ in~\eqref{scm_y}. Similarly, we have the function class $\mathcal{L}^{\tau}$ for the testing SCM $\mathcal{S}^{\tau}$. It is important to note that even though $\E_{l}[Y|X; U]$ achieves the minimum prediction error for $Y$ (as in~\eqref{eq:lmmse_U}), it may not be applicable for predicting $Y^{\tau}$ since $\mathcal{U}$ and  $\mathcal{U}^{\tau}$ may differ in general. And in fact, the prediction error of using $\E_{l}[Y|X; U]$ for $Y^{\tau}$ can be arbitrarily high as we do not restrict the forms of $\alpha(\cdot)$ and $\alpha^{\tau}(\cdot)$. In the next section, we show that this issue can be resolved via our invariance property. 



\section{Invariant Matching Property} 
\label{sec:invariance}

\subsection{One Motivating Example}
\label{subsec:example}
\begin{example}
\label{example1}
Consider $(Y,X^{\top},U) \triangleq (Y,X_{1},X_{2},X_{3},U)$ satisfying the following acyclic SCM (illustrated in Fig.~\ref{fig_toy}), \\
\begin{figure}[h]
\centering
 \vspace{-1em}
\includegraphics[width=0.40\linewidth]{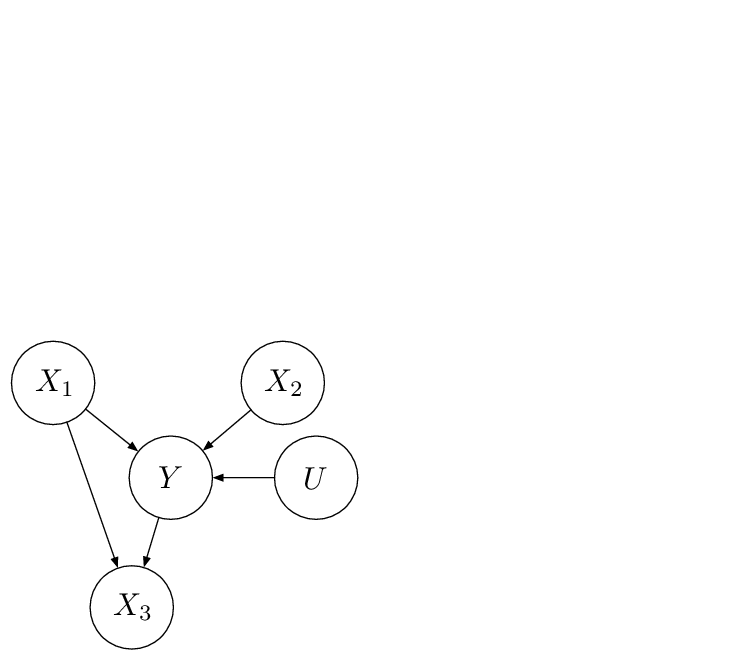}
 \vspace{-1em}
\label{fig_toy}
\caption{Directed acyclic graph $\mathcal{G}(\mathcal{S}_{\text{toy}})$. }
\end{figure}
\begin{align}
\mathcal{S}_{\text{toy}}:\label{scm_toy}
 \begin{cases}
    Y = a(U) X_1 + X_2 + N_{Y}\\
    X_3 = Y+ X_1 + N_3,
\end{cases}
\end{align}
where $U, X_1,X_2,N_{3}$, and $N_{Y}$ are jointly independent, and $ X_1,X_2,N_{Y}$, and $N_{3}$ are $\mathcal{N}(0,1)$-distributed. The testing SCM $\mathcal{S}_{\text{toy}}^{\tau}$ over $(Y^{\tau},X^{\tau}_{1},X^{\tau}_{2},X^{\tau}_{3},U^{\tau})$ can be defined similarly, where $X_{1}^{\tau}$ has a coefficient $a^{\tau}(U^{\tau})$. Since $(Y,X)$ is multivariate Gaussian given $U=u$, the MMSE estimator of $Y$ using $X$ given $U=u$ is
\begin{align}
\E[Y|X,U=u]  =& X^{\top}\left(\E[XX^{\top}|U=u]\right)^{-1}\E[XY|U=u]\nonumber\\
=& \frac{1}{2} (a(u)-1)X_{1} + \frac{1}{2}X_{2} + \frac{1}{2}X_{3}, \nonumber 
\end{align}
which implies $\E[Y|X,U] = \frac{1}{2} (a(U)-1)X_{1} + \frac{1}{2}X_{2} + \frac{1}{2}X_{3}$. Similarly, one can compute $\E[X_{3}|X_{1},X_{2},U] = (1+a(U))X_{1} + X_{2}$. Observe that $\E[Y|X_{1},X_{2},X_{3},U]$ and $\E[X_{3}|X_{1},X_{2},U]$ are two linear combinations of $\{a(U)X_{1}, X_{1}, X_{2}, X_{3}\}$; $a(U)X_{1}$ can not be linearly represented by $\{X_{1},X_{2},X_{3}\}$. Thus, there exists a deterministic linear relation
\begin{align}
\E[Y|X,U]    = \lambda\E[X_{3}|X_{1},X_{2},U] +\eta^{\top} X, \label{determin}
\end{align}
with unique coefficients $\lambda = 1/2$ and $\eta = (-1,0,1/2)^{\top}$ that \emph{do not depend on $U$}. Furthermore, since the right-hand side of~\eqref{determin} is a linear function of $\E[X_{3}|X_{1},X_{2},U]$ and $X$, and it equals to the MMSE estimator $\E[Y|X,U]$ among all functions of $X$ and $U$, we obtain an invariant relation
\begin{align}
  \E[Y|X,U] &= \E_l\biggl\{Y\,\biggl|\,\E[X_{3}|X_{1},X_{2},U],X\biggr\} \nonumber \\
  &=  \frac{1}{2}\E[X_{3}|X_{1},X_{2},U] -X_{1}+\frac{1}{2}X_{3} \label{toy_determin}
\end{align}
 for the training SCM. Since $\lambda$ and $\eta$ are not functions of $U$, they will remain invariant for the testing SCM. Note that $\E[X^{\tau}_{3}|X^{\tau}_{1},X^{\tau}_{2},U^{\tau}]$ is determined by the distribution of $(X^{\tau},U^{\tau})$. A prediction model like~\eqref{toy_determin} with invariant coefficients is often not unique when it exists. One can show that $\E[Y|X,U]= -\frac{3}{2}\E[X_{2}|X_{1},X_{3},U] - X_{1} + \frac{1}{2}X_{2}+X_{3}$, however, this does not hold for $\E[X_{1}|X_{2},X_{3},U]$. 

\end{example}
 \smallskip

\begin{remark}
In general, an invariance relation~\eqref{toy_determin} may not hold for the MMSE estimator $\E[Y|X,U]$ if $(X,Y)$ is not Gaussian when conditioning on $U=u$ for each $u \in \mathcal{U}$. In this work, we do not require Gaussianity, and we focus on the LMMSE estimator detailed in the next section.
\end{remark}

\subsection{Invariance Matching Property}

Our invariant matching property is motivated by the following observation: If $X$ includes any descendants of $Y$, then $\alpha(\cdot)$ (that may change in the unseen environments) will be passed on to the descendants. In other words, if $Y$ has at least one child, there will be certain dependency between the mechanism that generates $Y$ and certain statistical properties of $X$, which is also true for $Y^{\tau}$ and $X^{\tau}$. Thus, the change of the function $\alpha(\cdot)$ can be revealed by the changes of certain statistical properties of $X$. As illustrated in the motivating example, we have identify features of the form $\E_{l}[X_{k}|X_{S};U]$ to be useful for prediction. 

Formally, we say that a model $\mathcal{S}$ satisfies the \emph{invariant matching property} if there exists $k \in \{1,\ldots,d\}$ and $S \subseteq \{1,\ldots,d\}\setminus k$ such that
 \begin{align}
    \E_{l}[Y|X\,;\, U] &=\E_{l}\biggl\{Y\,\biggl|\,X,\E_l[X_{k}|X_{S}; U]\biggr\} \label{invar_train}  \\
    &= \lambda \E_l[X_{k}|X_{S}; U] + \eta^{\top}X \label{invar_coeff} ,
 \end{align}
where parameters $\lambda$ and $\eta$ do not depend on $U$, and the same holds for the testing SCM $\mathcal{S}^{\tau}$, i.e.,
\begin{align}
    \E_{l}[Y^{\tau}|X^{\tau}; U^{\tau}]  =  \lambda \E_l[X^{\tau}_{k}|X^{\tau}_{S}; U^{\tau}] + \eta^{\top}X^{\tau}. \nonumber
\end{align}
In general, due to the difference between $\alpha(\cdot)$ and $\alpha^{\tau}(\cdot)$,
\begin{equation}
      \E_{l}[Y|X=x; U=u] = \E_{l}[Y^{\tau}|X^{\tau}=x; U^{\tau}=u] \nonumber
\end{equation}
does not hold for $x \in \mathcal{X} \cup \mathcal{X}^{\tau}$ and $u \in \mathcal{U}\cup \mathcal{U}^{\tau}$ even if $U$ and $U^{\tau}$ are equal in distribution. By introducing some feature $\E_l[X_{k}|X_{S}; U]$, the invariant matching property bridges $\E_{l}[Y|X=x; U=u]$ and $\E_{l}[Y^{\tau}|X^{\tau}=x; U^{\tau}=u]$ (for the same $(x,u)$) with a linear relation that remains invariant across all observed and unseen environments. 



In our invariant matching property, note that~\eqref{invar_coeff} follows from~\eqref{invar_train} by the definition of linear MMSE. Now we show the other direction is also true in the following technical lemma.  
\smallskip
\begin{lemma}
\label{lem_invariance}
For some $k \in \{1,\ldots,d\}, S \subseteq \{1,\ldots,d\}\setminus k$, 
\begin{equation}
    \E_{l}[Y|X; U] 
    =  \E_{l}\biggl\{Y\,\biggl|\,X,\E_{l}[X_{k}|X_{S};U]\biggr\} \label{first_invariance}
\end{equation}
if and only if there exists $\lambda \in \mathbb{R}$ and $\eta \in \mathbb{R}^{d \times 1}$ such that
\begin{equation}
    \E_{l}[Y|X; U] = \lambda  \E_{l}[X_{k}|X_{S}; U] + \eta^{\top} X. \label{second_invariance}
\end{equation}
holds for all $u \in \mathcal{U}$. 
\end{lemma}

\subsection{Characterization of the Features}
\label{sec:poplutaiton}
An important fact about a feature of the form $\E_{l}[X_{k}|X_{S};U]$ is that it does not depend on $Y$, so the corresponding feature $\E_{l}[X^{\tau}_{k}|X^{\tau}_{S}; U^{\tau}]$ for the testing SCM will not depend on the unobservable variable $Y^{\tau}$. In other words, extracting the features only requires exploring the relations between the predictors while taking the target variables $Y$ and $Y^{\tau}$ as unobserved. The consequence of $Y$ being unobserved is that $(U,X_{1},\ldots,X_{d})^{\top}$ no longer follows a mixture of linear SCMs, but a mixture of linear models with a set of dependent noise variables. Specifically, when $Y$ is unobserved (or equivalently, substitute $Y$ in~\eqref{scm_y} into~\eqref{scm_x}), then the relations between the predictors are as follows, 
\begin{equation}
    X =\left( \gamma(\beta + \alpha(U))^{\top} + B \right)X+ \gamma \varepsilon_{Y}+\varepsilon_{X}, \label{only_X}
\end{equation}
where $\gamma \varepsilon_{Y}+\varepsilon_{X}$ is a vector of dependent random variables when $\gamma$ non-zero. Observe that the function $\alpha(\cdot)$ is captured by the relations of the predictors only if $\gamma$ is not a zero vector in~\eqref{only_X}, which brings up the following key assumption. 
\smallskip
\begin{assumption}
$Y$ has at least one child. \label{assump1}
\end{assumption}
\smallskip

The equivalent definition of our invariant matching property in~\eqref{second_invariance} allows for simpler evaluation through computation. In order to verify~\eqref{second_invariance} for a particular feature   $\E_{l}[X_{k}|X_{S}; U]$, we compute the LMMSE estimators $\E_{l}[X_{k}|X_{S}; U=u]$ and $\E_{l}[Y|X; U=u]$ and check whether there exists coefficients for~\eqref{second_invariance} to hold.  In the following theorem, we show that it holds for a wide class of $k \in \{1,\ldots,d\}$ and $S \subseteq\{1,\ldots,d\}\setminus k$. 

\medskip
\begin{theorem}
There exists $\lambda_{Y} \in \mathbb{R}$ and $\eta_{Y} \in \mathbb{R}^{d \times 1}$ such that
\begin{equation}
    \E_{l}[Y|X ; U=u] =  (\lambda_{Y}\alpha(u)+ \eta_{Y}) ^{\top} X \nonumber
\end{equation}
holds for every $u \in \mathcal{U}$. For each ${k} \in \{j: \alpha_{j}(u) = 0\}$ and $S \subseteq \{1,\ldots,d\} \setminus k$ such that $\{j: \alpha_{j}(u)\neq 0\} \subseteq S$, there exists $\lambda_{k,S} \in \mathbb{R}$ and $\eta_{k,S} \in \mathbb{R}^{d \times 1}$ such that
\begin{equation}
  \E_{l}[X_{k}|X_{S}; U=u] = (\lambda_{k,S}\alpha(u) + \eta_{k,S})^{\top} X \nonumber
\end{equation}
holds for every $u \in \mathcal{U}$. If $\lambda_{k,S}\ne 0$, then the relation
\begin{align}
        &\E_{l}[Y|X ; U=u] \nonumber\\
        =& \frac{\lambda_{Y}}{\lambda_{k,S}}  \E_{l}[X_{k,S}|X_{S}; U=u]  + \left(\eta_{Y}-\frac{\lambda_{Y}}{ \lambda_{k,S}}\eta_{k,S}\right) X, \nonumber
\end{align}
holds for every $u \in \mathcal{U}$. 
\end{theorem}
\smallskip

In this theorem, we provide a complete characterization of the invariant matching property~\eqref{invar_coeff} for  $\E_{l}[Y|X ; U=u]$ and $\E_{l}[X_{k}|X_{S}; U=u]$, which allows us to use $ \E_{l}[X_{k}|X_{S}; U]$ as a predictor for $Y$ (see Algorithm~\ref{alg1} for implementation details).

\begin{remark}
Observe that $\lambda_{k,S}$ is nonzero if and only if there exists different $u_{1},u_{2} \in \mathcal{U}$ such that
\begin{equation}
  \E_{l}[X_{k}|X_{S}; U=u_{1}] \neq  \E_{l}[X_{k}|X_{S}; U=u_{2}].
\end{equation}
 This is true in generic cases when $X_{k} \not\independent U | X_{S}$. Note that if Assumption~\ref{assump1} is not satisfied ($Y$ has no children), then $X_{k} \independent U | X_{S}$ since $U$ is a root node and has only one descendent $Y$.
\end{remark}

\section{Algorithm}

For each $k \in \{1,\ldots,d\}$, $S \subseteq \{1,\ldots,d\}\setminus k$, we estimate the following LMMSE estimator for the prediction of $Y$,
\begin{align}
  Y(k,S) &= \E_{l}\biggl\{Y\,\biggl|\,X,\E_l[X_{k}|X_{S}; U]\biggr\} \nonumber\\
  &= \lambda(k,S) \E_l[X_{k}|X_{S}; U] + \eta^{\top}(k,S)X.\label{lmmse_alg}
\end{align}
According to Lemma~\ref{lem_invariance}, a feature $\E_l[X_{k_{*}}|X_{S_{*}}; U]$ that satisfies the invariant matching property if and only if $Y(k_{*},S_{*})$ achieves the minimum prediction error for $Y$ among the prediction errors of all possible $Y(k,S)$'s. Since such feature is not unique in general (shown by the toy example in Section~\ref{subsec:example}), we do not choose the feature that leads to the lowest prediction error. Rather, we look for features with prediction errors below a certain threshold $\varepsilon$ (see the end of this section for the determination of $\varepsilon$). 

For the training SCM $\mathcal{S}$ and testing SCM $\mathcal{S}^{\tau}$, let $\mathcal{U} =  \{u_{1},\ldots,u_{p}\}$ and we denote $ \mathcal{U}^{\tau} \triangleq \mathcal{V} = \{v_{1},\ldots,v_{q}\}$ for simplicity of notation. For each $u_{i} \in \mathcal{U}$, we are given the \iid training data $\bs{X}^{u_{i}}\in  \mathbb{R}^{n(u_{i}) \times d}, \bs{Y}^{u_{i}} \in \mathbb{R}^{n(u_{i}) \times 1}$, and for each $v_{i} \in U^{\tau}$, we observe the \iid testing data $\bs{X}^{v_{i}} \in \mathbb{R}^{m(v_{i}) \times d}$.  Let $\bs{X} \in \mathbb{R}^{n \times d}$ with $n \triangleq \sum_{i=1}^{p}n(u_{i})$ denote the pooled data matrix of all $\bs{X}^{u_{i}},u_{i}\in \mathcal{U}$. Similarly, we define the pooled data matrices $\bs{Y} \in \mathbb{R}^{n \times 1}$ and $\bs{X}^{\tau} \in \mathbb{R}^{m \times d}$ with $m \triangleq \sum_{i=1}^{q}m(u_{i})$.

In Algorithm~\ref{alg1}, for $k \in \{1,\ldots,d\}, S \in \{1,\ldots,d\}\setminus k$, we adopt the OLS estimator to estimate the feature vectors
\begin{align}
    \hat{\E}_{l}[{\bs{X}}^{u_{i}}_{k}|\bs{X}^{u_{i}}_{S}] &= \bs{X}^{u_{i}}_{S}\left((\bs{X}^{u_{i}}_{S})^{\top}\bs{X}^{u_{i}}_{S}\right)^{-1}(\bs{X}^{u_{i}}_{S})^{\top}{\bs{X}}^{u_{i}}_{k}, \label{lmse_x}\\
\hat{\E}_{l}[{\bs{X}}^{v_{i}}_{k}|\bs{X}^{v_{i}}_{S}] &=\bs{X}^{v_{i}}_{S} \left((\bs{X}^{v_{i}}_{S})^{\top}\bs{X}^{v_{i}}_{S}\right)^{-1}(\bs{X}^{v_{i}}_{S})^{\top}{\bs{X}}^{v_{i}}_{k}, \label{lmse_x_t}
\end{align}
for the training data and the testing data, respectively.

Let $\bs{\tilde{X}}(k,S) \triangleq \left(\hat{\E}_{l}[\bs{X}_{k}|\bs{X}_{S}],\bs{X}\right) \in \mathbb{R}^{n \times (d+1)}$ denote the augmented design matrix. For feature selection on the training data, we compute the prediction residuals of $Y$ as follows
\begin{align}
\bs{R}(k,S) &= \bs{Y}- \bs{\hat{Y}}(k,S), \label{alg_eq}
\end{align}
with an estimate of $\bs{Y}(k,S)$ (the vector form of $Y(k,S)$) as
\begin{align}
\bs{\hat{Y}}(k,S)  & = \bs{\tilde{X}}(k,S)\left(\bs{\tilde{X}}^{\top}(k,S)\bs{\tilde{X}}(k,S)
\right)^{-1}\bs{\tilde{X}}^{\top}(k,S)\bs{Y}\nonumber\\
&\triangleq\bs{\tilde{X}}(k,S)\beta(k,S),  \nonumber
\end{align}
where the OLS estimator $\beta(k,S) \in \mathbb{R}^{d \times 1}$ can be reused for predicting $\bs{Y}^{\tau}$. That is, on the testing data we compute
\begin{equation}
    \bs{\hat{Y}}^{\tau}(k,S) = \bs{\tilde{X}}^{\tau}(k,S)\beta(k,S), \label{alg_predict}
\end{equation}
where $\bs{\tilde{X}}^{\tau}(k,S) \triangleq \left( \hat{\E}_{l}[\bs{X}^{\tau}_{k}|\bs{X}^{\tau}_{S}],\bs{X}^{\tau} \right) \in \mathbb{R}^{m \times (d+1)}$. 

\begin{algorithm}[H]
\caption{\label{alg1} Generalizable Prediction via Invariant Matching
}
	\begin{algorithmic}
\Procedure{Select Features on the Training data}{}
    	\For{$k \in \{1,\ldots,d\}$} 
    	    \For{$S \subseteq \{1,\ldots,d\}\setminus k$} \State{(i) Compute the feature vector $\hat{\E}_{l}[{\bs{X}}^{u_{i}}_{k}|\bs{X}^{u_{i}}_{S}]$ for}
    	    \State{each $u_{i}$ by~\eqref{lmse_x}, and combine the feature vectors} 
    	    \State{into one vector $\hat{\E}_{l}[\bs{X}_{k}|\bs{X}_{S}]$ }
    	    \State {(ii) Compute $\bs{R}(k,S)$ in~\eqref{alg_eq} and check whether} \State{  $||\bs{R}(k,S)||^{2}_{2} \leq \varepsilon$  }
    	    \EndFor
    	\EndFor
\EndProcedure	
\Procedure{Extract The Selected Features on the testing data }{}
\For{every $(k,S)$ such that $||\bs{R}(k,S)||^{2}_{2} \leq \varepsilon$}
\State{(i) Compute the feature vector $\hat{\E}_{l}[{\bs{X}}^{v_{i}}_{k}|\bs{X}^{v_{i}}_{S}]$ for each} \State{$v_{i}$ by~\eqref{lmse_x_t}, and combine the feature vectors into one} 
\State{vector $\hat{\E}_{l}[\bs{X}^{\tau}_{k}|\bs{X}^{\tau}_{S}]$}
\State{(ii) Predict $\bs{Y}^{\tau}$ using the feature $\hat{\E}_{l}[\bs{X}^{\tau}_{k}|\bs{X}^{\tau}_{S}]$ by}
\State{computing $\bs{\hat{Y}}^{\tau}(k,S)$ according to~\eqref{alg_predict}}
\EndFor
\EndProcedure	
\State{Output $\bs{\hat{Y}}^{\tau}$ as the average of all computed $\bs{\hat{Y}}^{\tau}(k,S)$}
	\end{algorithmic}
\end{algorithm}

We determine the parameter $\varepsilon$ using the residuals $\bs{R}(k,S)$ defined above. First, we run the first procedure in Algorithm~\ref{alg1} with a sufficiently large $\varepsilon$ to compute the training prediction error $||\bs{R}(k,S)||^{2}_{2}$ for all $(k,S)$'s. Then, we rank all the prediction errors, and set $\varepsilon$ be the $(100\alpha)\%$-quantile of the all prediction errors, where $\alpha$ controls the proportion of the features that will be selected. For the experiments in the next section, $\alpha$ is fixed to be $0.05$.  

\section{Experiments}
We compare our method with three baseline methods: Ordinary Least Squares (OLS), stabilized regression (SR)~\cite{pfister2021stabilizing}, and anchor regression (AR)~\cite{rothenhausler2021anchor}. For the anchor regression, we use a $5$-fold cross-validation procedure to select the hyper-parameter $\gamma $ from $\{0.2,0.4,\ldots,1\} \cup\{2,3,\ldots,5\}$.

 \begin{figure}[h]
\centering
\includegraphics[width=0.80\linewidth]{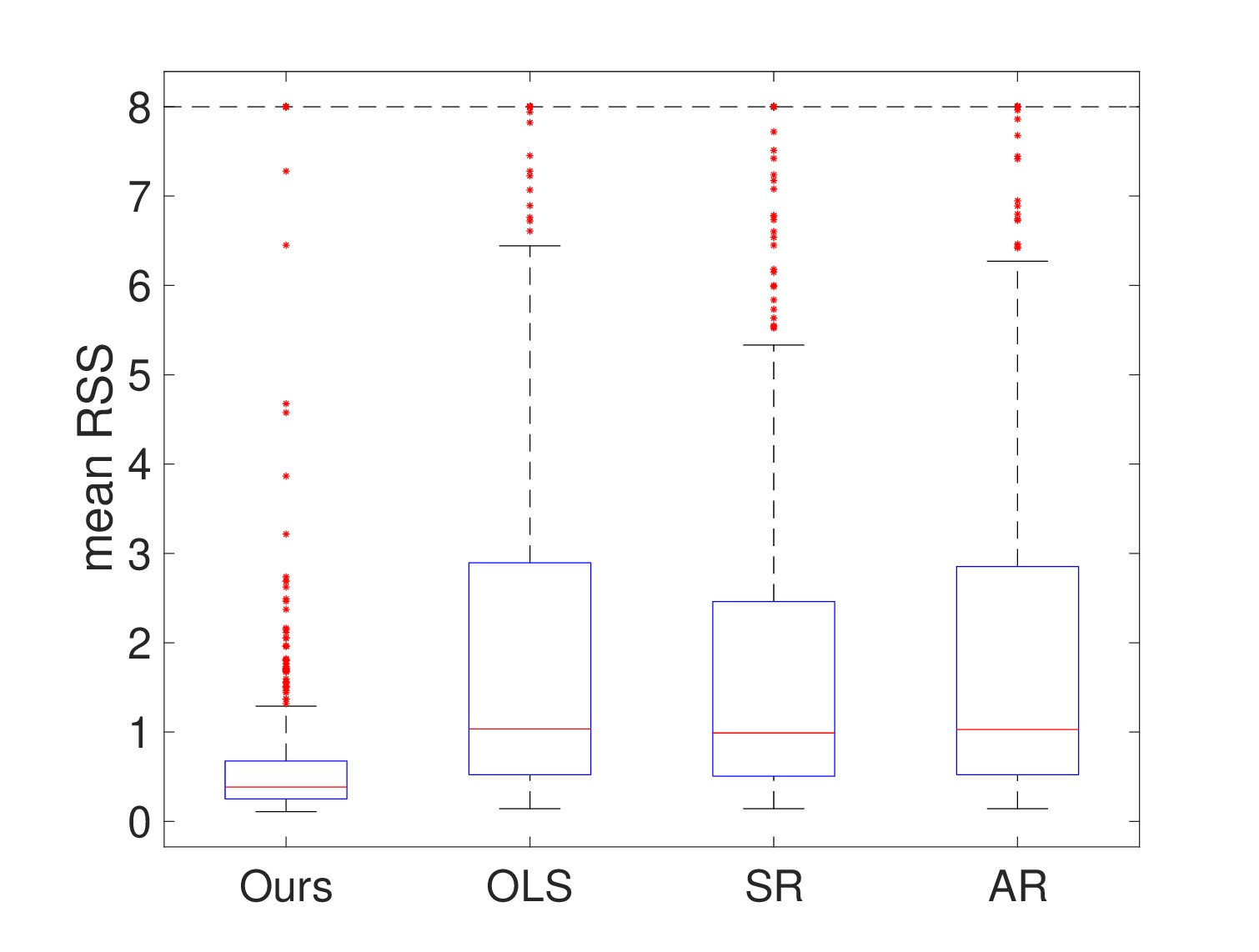}
 \vspace{-0.5em}
\caption{Experiment A. \label{fig1}
 } 
\end{figure}

\medskip
\noindent\underline{\textbf{Experiment A: Regular setting.}} 
\smallskip

We randomly simulate 500 models generated as follows. Consider the training SCM $\mathcal{S}$ with $10$ predictors and $\mathcal{U}= \{1,2,\ldots,5\}$. The acyclic graph $\mathcal{G}(\mathcal{S})$ is randomly generated with each edge existing with probability $0.5$. In the generated graph, we require $Y$ to have at least one parent and one child. For the parameters in the training SCM, all the coefficients in $\mathcal{S}$ that do not depend on $U$ are randomly sampled from $\U[-1.5,-0.5]\cup[0.5,1.5]$, and the noise variables $\varepsilon_{X}$ and $\varepsilon_{Y}$ are jointly independent standard normal random variables. For the intervention on $Y$, we choose $n_{p} \sim \U \{1,\ldots,|\text{PA}(Y)|\}$ of the parents of $Y$ to have a coefficient vector $\alpha(u)+\beta \in \mathbb{R}^{n_{p} \times 1}, u \in \mathcal{U}$. The coefficient vectors $\alpha(1),\alpha(2),\ldots$, $\alpha(5)$ are vectors of \iid random variables following  $\U[-2,2]$, and the elements of $\beta$ are sampled from $\U[-1.5,-0.5]\cup[0.5,1.5]$.

 The testing SCM $\mathcal{S}^{\tau}$ has the same graph and parameters as $\mathcal{S}$ except that $\mathcal{U}^{\tau} = \{6,7,\ldots,10\}$ and the new coefficient vectors $\alpha^{\tau}(6),\alpha^{\tau}(7),\ldots$, $\alpha^{\tau}(10)$ are drawn from \iid $\U[-10,10]$. For each $u \in \mathcal{U}$ or $u^{\tau} \in \mathcal{U}^{\tau}$, the sample size is 300. Overall, Fig.~\ref{fig1} shows our method outperforms all three baseline methods by having smaller median and variance for the mean residual sum of squares (RSS).

\begin{figure}[h]
\centering
\includegraphics[width=0.80\linewidth]{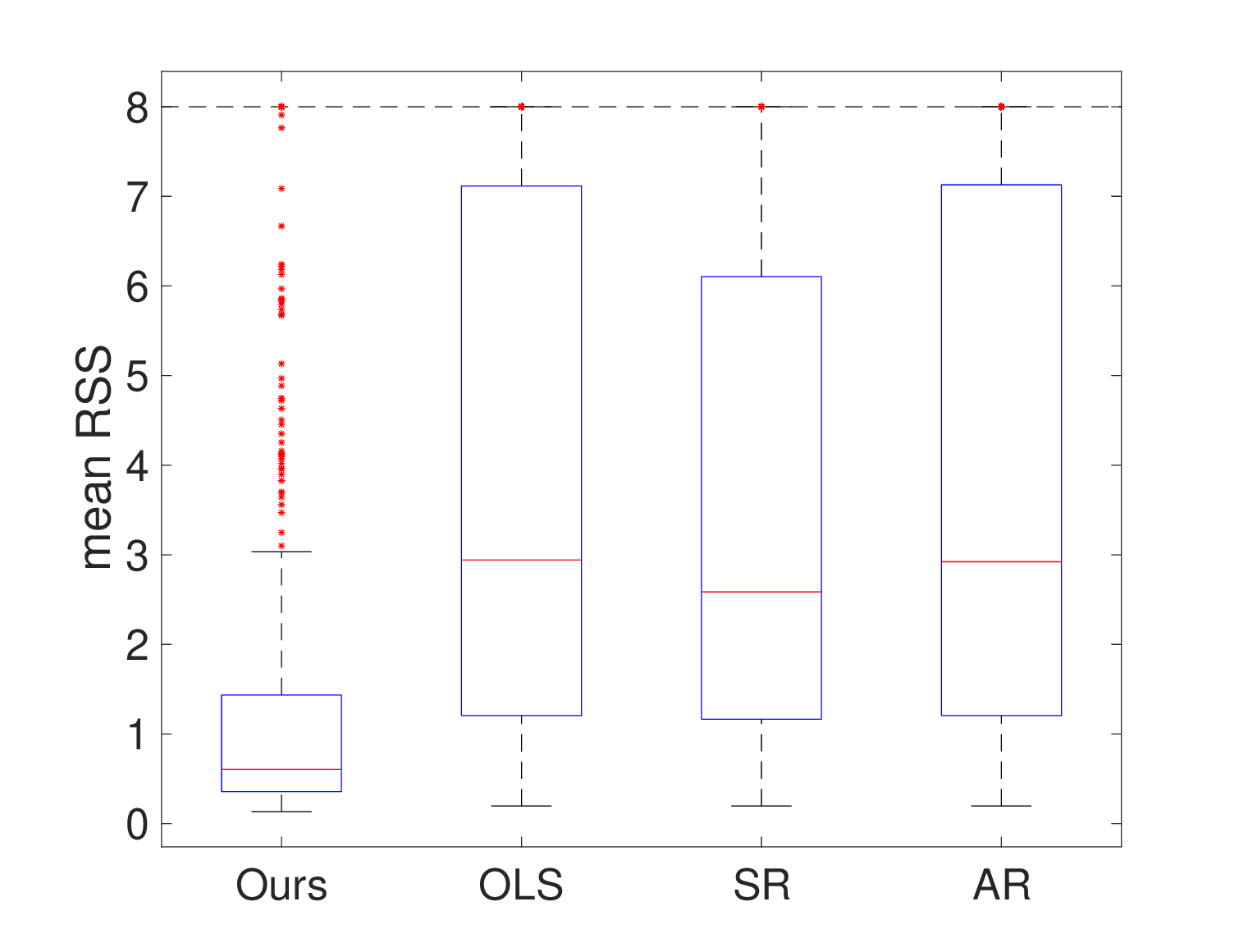}
 \vspace{-0.5em}
\caption{Experiment B-1. \label{fig2}
} 
\end{figure}

\begin{figure}[h]
\centering
\includegraphics[width=0.80\linewidth]{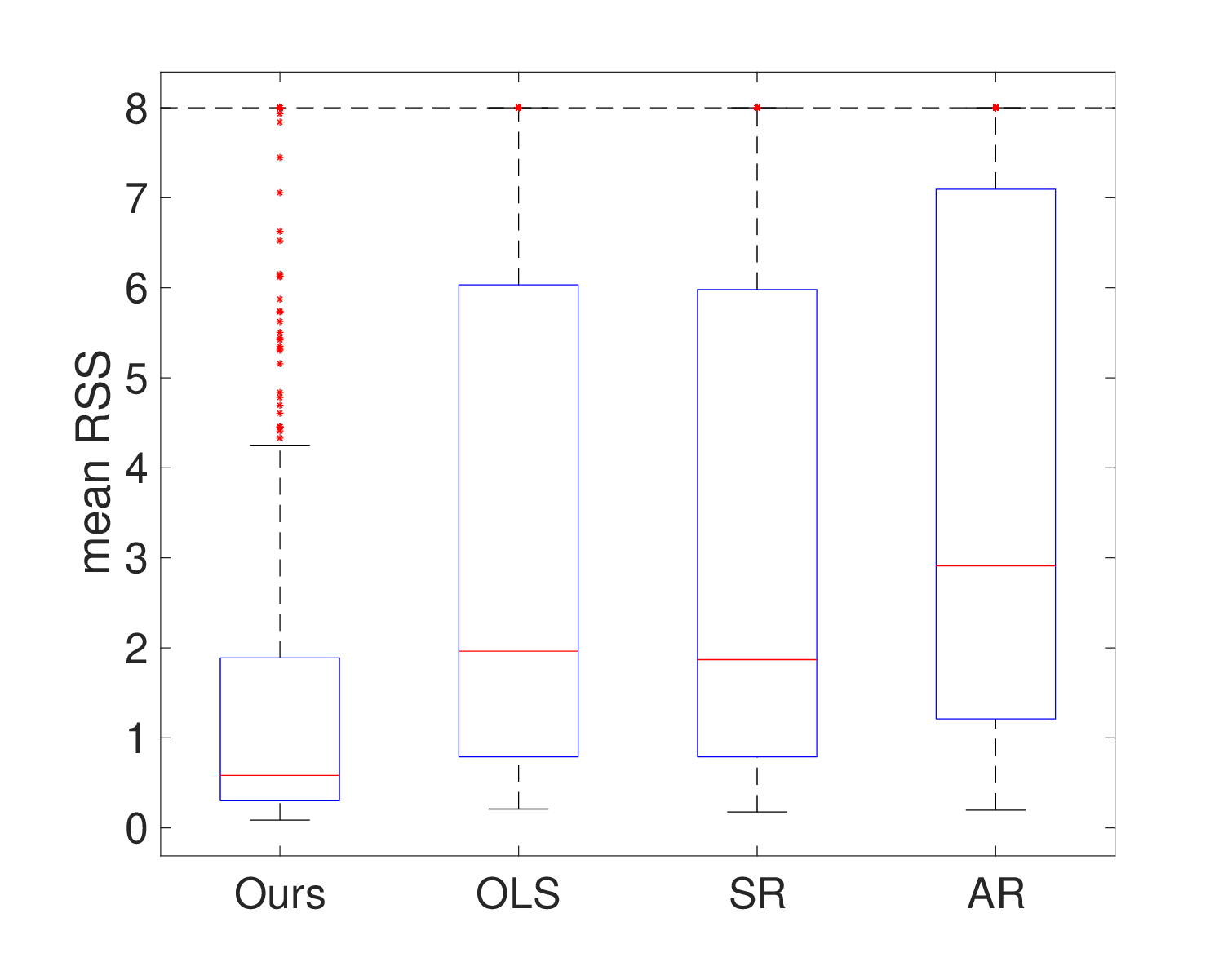}
 \vspace{-0.5em}
\label{fig3}
\caption{Experiment B-2. \label{fig3}
} 
\end{figure}

\medskip
\noindent\underline{\textbf{Experiment B: Reduced intervention on training data.}} 
\smallskip

We consider cases when interventions on the training data are reduced from that in Experiment A, while the data generating process for the testing data remains the same as before.

 \smallskip
\subsubsection{Smaller variation of coefficient vectors}
 The coefficient vectors $\alpha(1),\ldots$, $\alpha(5)$ are vectors of \iid entries according to $\U[-1,1]$, reducing the variation of the coefficient vector. 
 
  \smallskip
\subsubsection{Less number of environments}
 The support of the variable $U$ is reduced to $\mathcal{U} = \{1,2\}$. Accordingly, the sample size of the pooled training data is now $2*300 =600$.
 \smallskip
 
 In Fig.~\ref{fig2} and Fig.~\ref{fig3}, our method has a smaller median compared with the three baseline methods, while they have similar medians. Due to the averaging procedure over multiple prediction models in Algorithm~\ref{alg1}, our method has smaller variances than OLS and AR. The averaging procedure of SR fails since their assumption that $Y$ is not intervened is violated. Compared with Experiment~A, the median and variance of our method are slightly larger, but our method is less sensitive with respect to the reduced interventions in comparison with the baseline methods.

\balance
\bibliographystyle{IEEEtran}
\bibliography{ref.bib}


\end{document}